\documentstyle[preprint,aps,eqsecnum,epsfig]{revtex}

\def\beq{\begin{equation}}
\def\eeq{\end{equation}}
\def\eeq{\end{equation}}
\def\bea{\begin{eqnarray}}
\def\eea{\end{eqnarray}}
\def\bq{\begin{quote}}
\def\eq{\end{quote}}

\def\gtrsim{\mathrel{\mathpalette\vereq>}}

\makeatletter
\def\vereq#1#2{\lower3pt\vbox{\baselineskip1.5pt \lineskip1.5pt
\ialign{$\m@th#1\hfill##\hfil$\crcr#2\crcr\sim\crcr}}}
\makeatother
\begin{document}
\draft
{\tighten
\preprint{\vbox{\hbox{UCB-PTH-99/41}
		\hbox{LBNL-44254}
                \hbox{SLAC-PUB-8250}
                \hbox{September 1999}
                \hbox{}}}

\title{Split Fermions in Extra Dimensions and \\
Exponentially Small Cross-Sections at Future Colliders}

\author{Nima Arkani-Hamed,\,$^a$\thanks{\tt arkani@thsrv.lbl.gov} 
        Yuval Grossman\,$^b$\thanks{\tt yuval@slac.stanford.edu} and 
        Martin Schmaltz\,$^b$\thanks{\tt schmaltz@slac.stanford.edu}}
\address{ \vbox{\vskip 0.truecm}
  $^a$Department of Physics, University of California,
                 Berkeley, CA 94720\\
    and Theoretical Physics Group, LBNL, University of California,
                 Berkeley, CA 94720. \\ 
\vbox{\vskip 0.truecm}
  $^b$Stanford Linear Accelerator Center \\
  Stanford University, Stanford, CA 94309 }

\maketitle

\begin{abstract}%
We point out a dramatic new experimental signature for a class of theories with
extra dimensions, where quarks and leptons are localized at slightly separated
parallel ``walls'' whereas gauge and Higgs fields live in the bulk of the extra
dimensions.  The separation forbids direct local couplings between quarks and
leptons, allowing for an elegant solution to the proton decay problem.  We show
that scattering cross sections for collisions of fermions which are separated
in the extra dimensions vanish {\it exponentially} at energies high enough to
probe the separation distance. This is because the separation puts a lower
bound on the attainable impact parameter in the collision.  We present cross
sections for two body high energy scattering and estimate the power with which
future colliders can probe this scenario, finding sensitivity to inverse
fermion separations of order 10-70 TeV.  
\end{abstract}
} % end tighten
\newpage

%%%%%%%%%%%%%%%%%%%%%%
\section{Introduction}
%%%%%%%%%%%%%%%%%%%%%%

Any extension of the Standard Model (SM) with a low fundamental scale $M_*$
has to explain why it does not predict rapid proton decay through higher
dimensional operators suppressed by $M_*$. In the SM proton
decay is not a problem as the lowest-dimensional baryon number violating
operator is dimension six and is harmless when suppressed by the enormous
value of the Planck mass. However this becomes a serious problem
in theories which attempt to nullify the hierarchy between the Planck
scale and the weak scale by postulating that the fundamental scale
of gravity really is at or near a TeV, and that the apparent weakness of
gravity is due to a large extra-dimensional volume into which
the gravitational field can spread \cite{XD,AD,other,DDG,RS,More}.

One approach to this problem is to forbid proton decay
by postulating a new symmetry.
This symmetry would have to be gauged as gravitational
effects involving virtual black holes and worm holes violate global
symmetries and generate dangerous $M_*$-suppressed
proton decay operators. But anomaly cancellation conditions for gauge
symmetries make it difficult to find consistent theories. The
only known example is baryon triality \cite{Ross} which stabilizes
the proton but has baroque charge assignments. 

A different solution to the proton decay
problem which does not rely on symmetries but rather
exploits the new space in the extra dimensions was proposed in 
Ref. \cite{AS}.\footnote{The
solution proposed in
%by Dienes {\it et al.} 
\cite{DDG} does not stabilize the proton.  
Ref.\cite{DDG} proposes an orbifold $Z_2$ symmetry which does not allow
GUT gauge interactions to contribute to proton decay. 
However, this symmetry does not forbid any of the fatal higher dimensional
operators such as $QQQL$ which lead to disastrous proton decay.
These operators are the major problem for theories with a low fundamental
scale, even in the absence of GUT symmetries.}
The idea is to separate quarks from leptons in the extra dimensions.
Consider for example a model where the SM gauge and Higgs fields
live in the bulk of one extra compact dimension of radius TeV$^{-1}$
while the quarks and leptons are localized at different positions with
narrow wavefunctions in the extra dimension.\footnote{The gauge fields
may also be confined to a brane of thickness TeV$^{-1}$
in much larger extra dimensions. Then the fermions would be stuck
to thin parallel ``layers'' within the brane.} 
This separation of the fermion fields 
suppresses proton decay because direct couplings of quarks
to leptons are forbidden by five dimensional locality;
the proton decay rate is  exponentially suppressed by the
overlap of the quark and lepton wavefunctions.

At low energies ($E \ll$ TeV), experiments cannot resolve the size of the
extra dimension and its substructure. One observes
fermions coupled to the lightest modes of the gauge
fields with couplings exactly as in the SM.
Experiments at energies above TeV would discover a whole
tower of Kaluza Klein (KK) excitations of the gauge and Higgs fields,
proving that Higgs and gauge fields propagate in the bulk of an
extra dimension. Measurements of the couplings of the various KK fields
to the fermions can be used to map out the locations of the quarks and
leptons in the extra dimensions. But even at lower energies virtual KK mode
exchange leads to small deviations in precision measurements. For example,
as shown in \cite{AS}, quark lepton separation leads new contributions
to the prediction for atomic parity violation with the correct sign to
account for the measured deviations \cite{apv} from the SM value.

Even though we motivate the quark-lepton separation from proton decay
we note that fermion separation can be more general with all fermions
separated in the extra dimensions. In any realistic model the locations
of all the fermion fields are determined by potentials which depend
on the various parameters of the theory. Since the different SM fermion fields
have different gauge and Yukawa couplings we expect their potentials
to differ, leading to splittings in their positions.

In this paper we point out a dramatic and model independent experimental
signature of this scenario which follows simply from locality in the extra
dimensions: {\em At energies above a $TeV$, the large angle scattering
cross section for fermions which are separated in the extra dimensions
falls off exponentially with energy.}
This is easily understood from the fact that the fermion
separation in the extra dimensions implies a minimum impact parameter
of order TeV$^{-1}$. At energies corresponding to shorter distances
the large angle cross section falls off exponentially because the particles
``miss'' each other. The amplitude involves a Yukawa propagator for 
the exchanged gauge boson where the four dimensional momentum transfer 
acts as the mass in the exponential. More precisely, we find exponential
suppression in any $t$ and $u$ channel scattering of split fermions.
However, $s$ channel exchange is time-like, and therefore
the fermion separation in space does not force an exponential suppression.
Nevertheless, $s$ channel processes also lead to interesting signatures
as the interference of the SM amplitude with KK exchange diagrams
depends on the fermion separation.

The remainder of this paper is structured as follows: Section 2 reviews
the basic setup and explains how quark lepton separation suppresses
proton decay. In Section 3 we develop the necessary formulae to calculate
scattering cross sections in our framework. In Section 4 we apply the results
of section 3 to different physical systems (deep inelastic scattering,
$e^+e^-$ and $\mu^+\mu^-$ scattering) and show the reach and physics potential
of various colliders. Section 5 contains final discussion.

%%%%%%%%%%%%%%%%%%%%%%%%%
\section{Framework: extra dimensional geography}
%%%%%%%%%%%%%%%%%%%%%%%%%

In this section we describe our framework and review the ideas
which lead to it. Our starting point is the observation that
simple compactifications of higher dimensional theories
typically do not lead to chiral fermions.
The known mechanisms which do lead to chiral spectra usually
break translation invariance in the extra dimensions and
the chiral fermions are localized at special points in the compact space.
Examples include twisted sector fermions stuck at orbifold
fixed points in string theory, chiral states from intersecting D-branes,
or zero modes trapped to defects in field theory. Given that fermions
generically are localized at special points in the extra dimensions we
are motivated to consider the possibility of having different locations
for the different SM fermions. In such a scenario locality in the higher
dimensions forbids direct couplings between
fermions which live at different places. This suppression
of contact terms between fermions is very generic and leads to approximate
symmetries in the effective four dimensional theory. In our framework
the observed approximate global symmetries of the SM (such as baryon ($B$)
and lepton ($L$) number) are not accidental, they follow from non-trivial 
geography in the extra dimensions. 
The gauge and Higgs fields are necessarily bulk fields because they
need to couple to all the SM fermions. Gauge and Higgs field exchange
does generate non-local interactions but the effective operators obtained
in this way preserve $B$ and $L$ and cannot lead to proton decay. 

Let us discuss corrections to the above picture in detail. There are
two possible sources of interactions between quarks and leptons: 
direct local couplings (contact terms), or quarks and leptons
could both couple to a new de-localized ``bulk'' field which would act as
messenger and lead to couplings which are non-local in the extra
dimension.

Direct local interaction require
the wave functions of quarks and leptons to overlap. The resulting
effective four dimensional coupling is proportional to this overlap.
If, as in the model of \cite{AS}, the wave functions of the fermions
are Gaussian in the extra dimensions then the effective four
dimensional couplings are Gaussian in the distance between quarks and leptons.
A quark-lepton separation of 8 in units of the fermion
wave function's width leads to a factor $\sim \exp(-50)$ which
suppresses proton decay to safety.

What about non-local interactions via bulk messengers?
Generating proton decay requires a bulk messenger with $B$ and $L$
violating couplings. In addition, this messenger has to
be a fermion as the proton's fermion number has to be transferred
to the final state leptons. If the theory does contain a bulk fermion
with $B$ and $L$ violating couplings, we can estimate the strength of
the resulting effective proton decay operator. The relevant Feynman
diagram (in position space) involves the
Yukawa propagator of the messenger field from the quarks in
one location in the extra dimension to the leptons. For a messenger of mass
$M$ and a quark-lepton separation $d$ the propagator contains an
exponential $\exp(-Md)$. Thus, in order to avoid the proton decay bounds
we require that all bulk fermions with $B$ and $L$ violating
couplings be heavier than the inverse quark-lepton separation by
a factor of about 50. Note that even much lighter
bulk fermions can be harmless if $B-L$ is imposed as a gauge symmetry.
Then the messenger fermion also needs to carry the $B-L$ charge of the
proton in order to be dangerous.

While it will not be of central importance for this paper we
would like to mention a particularly satisfying picture for the
origin of the fermion separations in the context of fermion zero
modes stuck at defects:  
Assume that the SM is unified into $SO(10)$ in the 
five-dimensional theory at energies near $\sim 10$ TeV. Then splitting between
quarks and leptons at lower energies has a natural explanation if the
fermion fields' localization potential contains terms which couple to a GUT
symmetry breaking vacuum expectation value in the $B-L$ direction
\cite{AHS}.

We note that in addition to quark-lepton separation there may also
be splittings between the generations. The separation of left- and
right-handed components of the SM quarks and leptons could then explain
the hierarchies in the SM Yukawa couplings \cite{AS}. The separations
needed to produce realistic quark and lepton masses are in the range
$(0..5)$ in units of the wave-function width in the case of Gaussian
wave functions \cite{AS}. Explicit examples that reproduce the
observed femion masses were worked out in \cite{AMS}. 

Let us summarize the scales involved in the theory. The
lowest experimentally allowed radius is about
$(3\,$TeV$)^{-1}$ \cite{RW}. (In \cite{RW} no seperation was assumed. 
While their results do not directly apply to our case, the order of 
magnitude of the bounds should be the same.)
At energies above a few TeV the theory becomes effectively
higher dimensional, but we can continue to use a four dimensional
description by including KK excitations for the bulk gauge and Higgs fields.
The loop expansion parameter in this effective theory is
\beq
{g^2 N_{KK} \over 16 \pi^2},
\eeq
where $g$ stands for any of the SM gauge couplings and $N_{KK}$ is the
number of KK excitations contributing in the loop. Our perturbative
description of physics breaks down when this parameter is of order unity
which occurs for $N_{KK}\sim 100$ or $M_*\sim 100\,$TeV. The width of the
fermion wave functions in the extra dimension is more model dependent.
In the field theoretic construction of \cite{AS} it must be at least a
factor of 10 narrower than the separation in order to sufficiently
suppress proton decay.

It should be clear from this discussion that the scale of quark-lepton
separation is well below the scale where the theory becomes strongly
coupled, and where quantum gravity or stringy effects may become important.
The fermion separation serves as an energy cut-off and suppresses
incalculable high energy contributions from the unknown theory of
quantum gravity.

%%%%%%%%%%%%%%%%%%%%%%%%%%%%%%%%%%
\section{Scattering of fermions localized at different places}
%%%%%%%%%%%%%%%%%%%%%%%%%%%%%%%%%%

\subsection{One extra dimension}

Let us now imagine colliding fermions which are localized at two
different places in a circular extra dimension of radius $R$.
Motivated by the solution to the proton decay problem discussed above,
we will begin by considering the 
scattering of electrons on protons, although 
we can imagine more generally that any set of the $(Q,U^c,D^c,L,E^c)$ fields
are split in the extra dimensions; 
indeed  our most interesting experimental signatures will be for the case of 
separations in the lepton sector. 

In the context of our model there are three potentially relevant mass
scales for this collision: the momentum transfer of the $t$-channel
scattering $\sqrt{-t}$,
the inverse of the quark-lepton separation $d^{-1}$ which we take
to be of order of the inverse thickness $R^{-1}$ of 
the extra dimension, and the inverse
width of the fermion wave functions $\sigma^{-1}$. However, as discussed above
proton stability requires quarks to be well separated from leptons and we
will approximate the fermion wave functions by delta functions
for the calculation. At the end of this section we will compute the
corrections which arise from the finite width of the wave functions
and verify that they are negligible for practical purposes.

To calculate the scattering though intermediate bulk gauge fields
we can either choose to work with a five-dimensional propagator
directly or else add contributions from an infinite tower of KK excitations
in a four dimensional context. It is instructive to do it both ways.
The five dimensional propagator in momentum space is $(t-p_5^2-m^2)^{-1}$
where we separated out the five dimensional momentum transfer $p_5$.
As we are interested in propagation between definite positions in
the fifth dimension it is convenient to Fourier transform in the
fifth coordinate
\beq
P_d(t)= \sum_{n=-\infty}^{\infty} {e^{ind/R}
    \over t-({n/R})^2-m^2}\ ,
\label{fiveprop}
\eeq  
where $d=x_q-x_l$ and $x_f$ is the location of fermion $f$ in the 
extra dimension. The Fourier transform is a sum and not an 
integral since momenta in the fifth coordinate
are quantized in units of $1/R$.

This propagator can also be understood in the four dimensional (4d) language
as arising from exchange of the 4d gauge boson and its infinite tower of KK
excitations. To see this expand the KK excitations of the gauge field
in plane waves, $\exp(i n x_5/R)$. 
Each of these KK modes has a four dimensional propagator
$(t-({n/R})^2- m^2)^{-1}$. 
Furthermore, the couplings to the fermions differ
for the various KK gauge bosons. They follow from expanding the
five dimensional action
%\bea
% \int dx_5\, d^4{\bf x}\ \delta(x_5-x_f)\;g \;
%   \overline \Psi({\bf x}) \not\hskip-.12cm{A}({\bf x},x_5)\,\Psi({\bf x}) =
%& & \\   
%\int d^4{\bf x} \sum_n\ g\, e^{i n x_f/R}\
%   \overline \Psi({\bf x}) \not\hskip-.12cm{A}^n({\bf x})\,\Psi({\bf x}) \,.
%& & \nonumber
%\eea
\beq
 \int dx_5\, d^4{\bf x}\ \delta(x_5-x_f)\;g \;
   \overline \Psi({\bf x}) \not\hskip-.12cm{A}({\bf x},x_5)\,\Psi({\bf x}) =
   \int d^4{\bf x} \sum_n\ g\, e^{i n x_f/R}\
   \overline \Psi({\bf x}) \not\hskip-.12cm{A}^n({\bf x})\,\Psi({\bf x}) \,.
\eeq
Thus the modified couplings are $g_n=g e^{i nx_f/R}$.
We can now write the ``KK-tower propagator'' which
is a sum over the propagators of the KK modes, including phase
factors from the modified couplings. The final expression is the same as
eq.~(\ref{fiveprop}).

This propagator can be simplified by performing the sum. To this end one
rewrites it as a contour integral with a cigar-shaped contour that
encircles the real axis and then deforming the contour
\beq
P_d(t)= \oint {dn\over 2\pi i}\ {\pi \over \sin(\pi n)}\
  {e^{i n(d/R-\pi)} \over t-(n/R)^2-m^2}\,. 
\eeq
Performing the integral we find
\beq
P_d(t)= - {\pi R\over \sqrt{-t+m^2}}\ 
  {\cosh[(d-\pi R)\sqrt{-t+m^2}]\over \sinh[\pi R\sqrt{-t+m^2}]}\,.
\label{towerprop}
\eeq
The Feynman rules for diagrams involving exchange of bulk gauge fields
are now identical to the usual four dimensional SM Feynman rules
except for the replacement of 4d gauge boson propagators by the corresponding
5d propagators. Before we proceed with calculating cross sections
we note a few properties of the propagator we just found.

It is easy to understand the two limits
$\sqrt{-t}\gg R^{-1}$ and $\sqrt{-t}\ll R^{-1}$. In the former
case we obtain 
\beq
P_d(t) \simeq - {\pi R\over \sqrt{-t}}\ e^{-\sqrt{-t}\,d}\,,
\eeq
which vanishes exponentially with the momentum transfer in the process
as we anticipated from five dimensional locality.
In the limit of small momentum transfer we obtain
\beq \label{small-t}
P_d(t) \simeq {1\over t-m^2} - 
       R^2\left(\frac{d^2}{2R^2}-{d \pi \over R}+\frac{\pi^2}{3}\right)\,,
\eeq
which is the four dimensional $t$-channel propagator
plus a correction term whose sign and magnitude depends on the
fermion separation. For small separation $d < \pi R\,(1-1/\sqrt{3})$ the 
correction enhances the magnitude of the amplitude, while for larger separation
it reduces it.  

It is also instructive to expand the propagator in exponentials (ignoring
the mass $m$)
\beq \label{expand}
P_d(t)=- {\pi R\over \sqrt{-t}} 
       \left( e^{-\sqrt{-t}d}+e^{\sqrt{-t}(d-2\pi R)}\right)
       \left( 1+e^{-\sqrt{-t}2\pi R}+e^{-\sqrt{-t}4\pi R}+
              \dots\right)\,,
\eeq
which can be understood as a sum of contributions from five
dimensional propagators. The two terms in the first parenthesis
correspond to propagation from $x_q$ to $x_l$ in clockwise and 
counter--clockwise directions, and the series in the other parenthesis adds
the possibility of also propagating an arbitrary number of times around the
circle. 

For later use we note that the expression for the $u$-channel KK-tower
propagator $P_d(u)$ is identical to eq.~(\ref{towerprop}) with the obvious
replacement $t\rightarrow u$, and $P_d(s)$ is obtained by analytic
continuation
\beq
P_d(s)= {\pi R\over \sqrt{s-m^2}}
  {\cos[(d-\pi R)\sqrt{s-m^2}]\over \sin[R\pi\sqrt{s-m^2}]}\ .
\label{schannel}
\eeq
The poles at $\sqrt{s-m^2}=n/R$ are not physical and can be avoided by including
a finite width $\Gamma$.
Note also that for $s \ll R^{-2}$ (but $s > m^2$) the relative sign between
the SM propagator and it's first correction is opposite to the corresponding
sign in the $t$ channel exchange case. Namely for small (large) separation the 
amplitude is smaller (larger) than the SM one.

Armed with this propagator it is easy to evaluate any KK boson
exchange diagram in terms of its SM counterpart. For example, a pure
$t$ channel exchange diagram becomes
\beq
{\cal M} =  (t-m^2) P_d(t) \times {\cal M}\big|_{SM} \ ,
\eeq
where ${\cal M}\big|_{SM}$ is the SM amplitude and the factor 
$(t-m^2) P_d(t)$
replaces the SM gauge boson propagator $1/(t-m^2)$ by the 5d
propagator $P_d(t)$. 

We now compute $P_d^{fw}(t)$, the propagator between fermions
which have a finite width in the extra dimension. The result is most
easily obtained by integrating
the propagator eq.~(\ref{towerprop}) over the wave functions of the initial
and final fermions
\beq \label{fw}
P_d^{fw}(t) =
\int dy dy' |f_q(y)|^2 P_{|y-y'|}(t) |f_l(y')|^2\,,
\eeq
where $f_q$ ($f_l$) is the quark (lepton) wave function.
For demonstration, we perform the integrations for the special case
of Gaussian wave functions
\beq \label{wave}
f_f(y) = {1 \over \pi^{1/4} \sigma^{1/2}} e^{-(y-x_f)^2/(2 \sigma^2)}\,,
\eeq
as in the model of \cite{AS}. We assume that the wave functions
are narrow compared to their separation and have common width.
We present below the result in two relevant limits. 
In both cases we assume
$\sqrt{-t} \gg R^{-1}$ (and therefore also neglect $m$).
In the first case, in an intermediate momentum regime we find
\beq
P_d^{fw}(t) =  e^{-t\sigma^2/2} \; P_d(t) 
\qquad {\rm for} \qquad \sqrt{-t}\ll d/\sigma^2\,.
\eeq
Not surprisingly, the amplitude is still exponentially suppressed, but
it is enhanced relative to the delta function approximation by a factor
which is significant for momenta large compared to the inverse width.
For much larger $t$ we find
\beq
P_d^{fw}(t) = 
  {-\sqrt{2\pi}R\over \sigma \,t}\, e^{-d^2/(2\sigma^2)}
                \qquad {\rm for} \qquad  \sqrt{-t}\gg d/\sigma^2 \,.
\eeq
In that limit the
scattering is dominated by direct local scattering through the small
but non-vanishing overlap of the fermion wave functions. 
The propagator has the normal 4d momentum dependence but the coupling is
suppressed by the exponentially small wave function overlap.
Since the energies attainable at upcoming colliders do not allow us to probe
distances shorter than the fermion wave function width the corrections
to the propagator of eq.~(\ref{towerprop}) can be ignored
for all practical purposes.

\subsection{$n$ extra dimensions}

In the case of $n>1$ extra dimensions of equal radius $R$,
a straightforward extension of
the above tells us that the propagator is that of the Yukawa propagator in 
$n$ (compact) dimensions. Let the separation be a vector $d_i$.
If $d \equiv |d_i| \ll R$ and $\sqrt{-t}\gg 1/R$, then the effects of the
compactness of the space are negligible and we find
the KK-tower propagator by a simple Fourier transform of the momentum space
propagator
\beq
P^{0}_{d_i}(t)=
          \int_{-\infty}^\infty d^np \ {e^{i d_i p_i /R}\over t-p_i^2/R^2}\ .
\eeq
The result is just the volume $(2 \pi R)^n$ times the 
Yukawa potential in the $n$ transverse dimensions, 
with mass $\sqrt{-t}$
\begin{equation} \label{bes}
P^{0}_{d_i}(t) = -\frac{(2 \pi R)^n}{(2 \pi)^{n/2}} 
\left(\frac{\sqrt{-t}}{d}\right)
^{(n-2)/2} K_{(n-2)/2} (\sqrt{-t} d)\,,
\end{equation}
where $K_p$ is the modified Bessel function.
For large $\sqrt{-t} d$, we use the large argument limit of the
Bessel function to see the exponential suppression explicitly
\begin{equation}
P^0_{d_i}(t) \rightarrow -\left({2 \pi R^2 \sqrt{-t}\over d}\right)^{(n-1)/2}\
{\pi R \over \sqrt{-t}}\ e^{-\sqrt{-t}d} \,.
\end{equation}
Including the effects of the finite size $R$ of the dimensions is 
easily done using the method of images,
\begin{equation}
P_{d_i}(t) = \sum_{k_i=-\infty}^{\infty} P^{0}_{d_i + 2 \pi k_i R} (t) \,,
\end{equation}
generalizing eq.~(\ref{expand}).
While this sum is not given by a simple closed form expression 
as in the case $n=1$, for all practical purposes only the first few 
images make a significant contribution.

There is an important feature for the case of two or 
more extra dimensions that deserves comment here. 
For unseparated fermions, the sum over tree-level exchange 
of KK gauge bosons is found to be 
UV divergent; the relevant sum is of the form
\begin{equation}
\sum_{n_i} \frac{1}{t - (n_i/R)^2} \sim R^n \int d^n k\frac{1}{t - k^2}\,,
\label{UVD}
\end{equation}
which is clearly UV divergent 
for $n \geq 2$, reflecting the singularity of the Yukawa potential
at short distances in two or more dimensions. 
This is usually dealt with by cutting the sum off at the 
fundamental scale $M_*$, but there is considerable uncertainty 
in doing this \cite{T}. It is easy to see 
that when the gauge boson exchange is between fermions 
separated in the extra dimensions, the 
separation acts as a natural cutoff and allows an unambiguous 
result to be obtained. The result
is just given by replacing the SM propagator with 
$P_{d_i}$, which is manifestly finite. The
usual UV divergence is seen in the singularity of 
$P_{d_i}$ as $d_i \to 0$.

Note that even in the absence of fermion 
separation, the width of the fermion wave function acts as a 
natural UV cutoff. Indeed, the integrand 
of eq.~(\ref{UVD}) should be multiplied by fermion wave functions 
in the higher-dimensional momentum space.
As it stands, eq.~(\ref{UVD}) corresponds to delta function 
wave functions in position space. Replacing the delta functions
with Gaussians of width $\sigma$ cuts off the UV 
divergence of eq.~(\ref{UVD}) at momenta of order $\sigma^{-1}$.
Explicitly we calculate the leading term for small $\sigma$ (and $d=0$)
using eqs.~(\ref{fw}), (\ref{bes}) and (\ref{wave})
%
%We can explicitly calculate the $d=0$ integral for fermions with finite width
%wave functions. Using eqs.~(\ref{fw}) and (\ref{bes}) and assuming wave 
%functions as in (\ref{wave}) in each of the compact dimensions, 
%the leading term for small $\sigma$ is given by
\bea
P^{fw}_0(t) \approx -{(2 \pi)^{n/2} \over n-2} {R^n \over \sigma^{n-2}}
& \qquad & n>2 \\
P^{fw}_0(t) \approx 2 \pi R^2 \log(\sqrt{-t} \sigma)
& \qquad & n=2 \,.\nonumber
\eea
These expressions are similar to the hard cut off results \cite{HLZ},
with $\sigma^{-1}$ playing the role of the cut off scale $M_S$.

%%%%%%%%%%%%%%%%%%%%%%%%%%%%%%%%%%%%%%%%%%
\section{Collider signatures}
%%%%%%%%%%%%%%%%%%%%%%%%%%%%%%%%%%%%%%%%%%
Having calculated the 5d propagator, the calculation of 
differential cross sections is a simple generalization of SM
results. The general SM results can be found in Ref. \cite{book}. 
To compute the differential cross section for deep inelastic scattering we
sum over contributions from neutral current exchange (photon and $Z$ plus
KK towers)\footnote{In the formulae 
in this section we neglect $m_Z$. It is easy to 
reintroduce it, and in our numerical plots we keep it.}
between the electron and all partons of the proton. Happily,
each term in the sum is simply equal to the SM term times $t P_d(t)$
which can be factored so that our final expression for the 
differential cross section of deep inelastic scattering becomes
\beq \label{ep} 
r_\sigma^t \equiv {{d\sigma / dt} \over
{d\sigma / dt} \left|_{\rm SM} \right. }
=  \left|t\, P_d(t)\right|^2 \,,
\eeq
where $P_d(t)$ is given in eq.~(\ref{towerprop}) and $t$ is the
measured difference between initial and final electron momentum
squared.
The effect of the KK tower would be seen as a dramatic
reduction of the cross section at large $-t=Q^2$.
To illustrate this point
in Fig. 1 we plot the ratio $r_\sigma^t$ of eq.~(\ref{ep}) as a function
of $t$ for $R=1\ TeV^{-1}$ and representative values of $d$.  

While an exponential suppression of the cross section would be an 
unambiguous signal of fermion separation in the extra dimension, we
can still probe $d$ if a small deviation of $r_\sigma$
from unity is found. The sensitivity can estimated from
eq.~(\ref{small-t}). Assuming maximum separation, $d=\pi R$, 
there is a reduction in the cross-section 
($r_\sigma^t < 1$), and we obtain
a sensitivity 
\beq \label{t-sens}
R \le \sqrt{3 \, \Delta r_\sigma^a\over \pi^2 Q^2}\,,
\eeq
where $\Delta r_\sigma^a$ is the combined theoretical and experimental error
on $r_\sigma^a$. For $d=0$ one should find $r_\sigma^t>1$ with 
a factor of $\sqrt{2}$ higher sensitivity.
At HERA, which is the only $e-p$ machine at present, we have
$\Delta r_\sigma^t$ at the few percent level. Thus, we
cannot obtain a strong bound from the HERA data.
In the future a more energetic machine may be built.
In the most optimistic scenario that is being discussed we may expect
a machine with $\Delta r_\sigma^t \approx 10\%$ at a maximum
$Q^2 \approx (4\,{\rm TeV})^2$ which will
be able to probe down to $R \approx (18\,$TeV$)^{-1}$.

Let us now switch gears and consider the predictions of our model for
high energy $e^+e^-$ or $\mu^+\mu^-$ machines. The
doublet and singlet components of the charged leptons may be split
by a distance $d$ in the extra dimensions. This would naturally
suppress the Yukawa couplings of the leptons and might be the origin
of the hierarchy $m_e/m_{top}$ \cite{AS}. In this case the wave
functions of the fermions cannot be arbitrarily narrow as the Yukawa
coupling is proportional to the overlap of the wave functions of the
doublet and singlet fermion. The finite width of the wave functions
ultimately cuts off the exponential suppression of $t$-channel scattering
amplitudes as discussed at the end of the previous section. This cut-off
is somewhat model-dependent as it depends on the
shape of the fermion wave functions. But if the separation of left and right
handed fields is responsible for at least part of the suppression of the
muon and electron Yukawa couplings then we can safely ignore the
finite width of the wave functions at energies relevant to experiments.

Again, to obtain any amplitude, we simply replace all SM gauge boson
propagators by their corresponding 5d propagators eqs.
(\ref{towerprop}) and (\ref{schannel}). If a given
cross section has only contributions in one channel ($s$, $t$ or $u$),
then it is given by the SM cross section multiplied by the
corresponding ratio of propagators as in the case of electron proton
collisions. A particularly clean measurement of $d$ would be possible
at a lepton collider with polarizable beams, as we could study
$l_L^+ l_R^- \longrightarrow l_L^+ l_R^-$ to isolate $t$ channel exchange.
In that case the deviation from the SM predictions is given by 
eq.~(\ref{ep}). 

We can get more information by 
combining the above with the processes
$e^+_N e^-_N \to \mu^+_N \mu^-_N$ ($N=L$ or $R$). 
(The same considerations also
apply to scattering into quark pairs, but this case is 
more difficult to study experimentally.)
This process is a pure
$s$ channel between {\it unseparated} fermions so that
\beq \label{ee-pol}
r_\sigma^{sN} \equiv {{d\sigma / dt} \over
{d\sigma / dt} \left|_{\rm SM} \right. }
=  \left|s\, P_0(s)\right|^2 \,.
\eeq
For $\sqrt{s}$ small
compared to the inverse size of the extra dimension
the cross section is reduced independently of $d$.
An extra dimensional theory without fermion separation predicts 
$r_\sigma^{sN} < 1$ and $r_\sigma^{t} > 1$.
Thus, a measurement of $r_\sigma^{sN} < 1$ together with $r_\sigma^t < 1$
would be evidence for fermion separation in the extra dimension.

Another interesting probe of $d$ using $s$ channel has been 
suggested recently \cite{T}.
Suppose that the first KK mode has been produced and its 
mass $1/R$ measured.
The case of $d=0$ can be distinguished from $d \neq 0$ by 
looking at the cross-section at lower energies. 
In particular, for $d=0$, the first KK exchange exactly cancels 
the SM amplitude at 
$\sqrt{s} = 1/(\sqrt{2}R)$, whereas for $d \neq 0$ the cross-section can 
still be large. 
Therefore, a beam scan at energies beneath the first resonance can 
be an efficient probe of $d$.
 
Even if beam polarization is not available, one can still 
probe the nature of the extra dimensions by looking at several processes
and using angular information. First consider an unpolarized
$e^+ e^- \to \ell^+ \ell^-$ scattering. (The same holds for incoming muons.)
We get the tree level cross section
\beq \label{ee-un}
{d\sigma \over dt} = {\pi \alpha^2 \over s^2} \left[
   \left(1+{1\over16\sin^4\theta_w}\right) 
{u^2 (P_0(s)+P_0(t))^2 \over \cos^4\theta_w}
   + {t^2P_d^2(s)+s^2P_d^2(t) \over 2 \cos^4\theta_w} \right]\,.
\eeq
When $\ell=e$ both $s$ and $t$ channels are possible, while for $\ell \ne e$
only the $s$ channel is present, and in the above formula one should set 
$P_d(t) = P_0(t) = 0$. We also define, as before, the ratio of the 5d 
cross section to the
SM one as $r_\sigma^{s}$ ($r_\sigma^{st}$) for the 
$e^+ e^- \to \mu^+ \mu^-$ ($e^+ e^- \to e^+ e^-$) reaction.
In Figs.~2 and 3, we presented
$r_\sigma^{s}$  and $r_\sigma^{st}$ as a function of the scattering angle.
As we can see, the cross sections depend in 
a non trivial way on the separation. This is
because the helicity changing amplitude depends on $d$, while the helicity
conserving one does not. 
By looking at angular distributions, one can separate the different 
contributions, and extract both $R$ and $d$. 

Another interesting collider mode which allows a very clean measurement
of fermion separations is $e^-e^-$ scattering. The advantage of the $e^-e^-$
mode is that both beams can be polarized to a high degree which
allows for a clean separation of the interesting $t$ and $u$ channels from
$s$ channel.
We find for $e_L^- e_R^-$ scattering to $e^- e^-$ (summed over final
polarizations)
\beq \label{emin-emin}
r_\sigma^{tu} \equiv {{d\sigma / dt} \over
{d\sigma / dt} \left|_{\rm SM} \right.} =
{u^2 |P_d(t)|^2 + t^2 |P_d(u)|^2 \over
             u^2/t^2 + t^2/u^2}\,.
\eeq

Last, we estimate the sensitivity of lepton colliders. 
Assuming $\Delta r_\sigma \approx 1\%$ and using eq.~(\ref{t-sens})
we conclude that we get sensitivity  
down to $R \approx (27\,$TeV$)^{-1}$ at a $1.5\,$TeV linear collider
and $R \approx (72\,$TeV$)^{-1}$ at a $4\,$TeV muon collider.

A hadron machine could also be used to probe extra dimensional separations.
Here, the situation is somewhat more complicated as there are many
subprocesses that contribute, the theoretical predictions are more uncertain
and the experimental situation is more complicated. However, the higher energy
of the hadron machine compensates for these drawbacks.

One possible probe is to look into dijet production, in particular, for 
high $p_T$ jets. This process occurs via $qq$, $q \bar q$ and $gg$ scattering
that occurs via $s$, $t$ and $u$ channels.
In our framework the first two will be modified in a way similar to
what we described for the leptons. In general, the invariant mass of the
two jets can be measured and thus one can find $\hat s$, the parton center
of mass of the event. Combining it with the angular information one
can determine both $s$ and $t$ for each event. 
This double differential cross section is sensitive to the size of the extra
dimension and the fermion separation. 
Another possible probe of our scenario is Drell-Yan processes.
Here, while one has less statistics, the accuracy is higher. 
In contrast to the dijet case, this is a pure $s$ channel.
Of course, for both of these cases a more detailed study needs
be done to see exactly what kind of sensitivity is attainable. 
Assuming $\Delta r_\sigma \approx 10\%$ 
at a maximum $Q^2 \approx (7\,{\rm TeV})^2$ and using eq.~(\ref{t-sens})
we estimate that one will be able to probe 
down to $R \approx (40\,$TeV$)^{-1}$.

We have so far contented ourselves to putting limits on the model, 
in some cases noting that the difference between extra 
dimensional models with and without fermion separation could be 
resolved. It is more exciting to 
consider how large a positive signal for exponentially dropping 
cross-sections could reasonably be
expected at future colliders. The direct limits from 
searching for the KK gauge bosons (and
$Z'$ searches) imply $1/R \geq 800\,$GeV. On the other hand, 
precision electroweak bounds on 
higher-dimensional operators generated by KK exchange place a far more 
stringent limit $1/R \gtrsim 3\,$TeV
\cite{RW}. If we take these precision bounds seriously, 
then a $1.5\,$TeV NLC could still observe a drop 
in the cross-section by as much as a factor of 2 for backscattering. 
However a $4\,$TeV muon collider could
see a reduction by as much as a factor of 60. More optimistically, 
we can imagine that there are extra states in the bulk  
whose exchange modifies the precision electroweak analysis. 
If these bounds are ignored,
the direct bounds are weak enough that spectacular drops 
in the cross-section can be observed, by as much as a 
factor of $1000$ at a $1.5\,$TeV NLC.

%%%%%%%%%%%%%%%%%%%%%%%%%%%%%%%%%
\section{Discussion}
%%%%%%%%%%%%%%%%%%%%%%%%%%%%%%%%%
Our signal displays a remarkable fact about scenarios with fermions split in
the extra dimensions. Traditionally, when fermion fields are either 
delocalized in the extra dimensions or when they are localized without any 
splitting, at energies above the compactification scale all the amplitudes 
grow faster than in 4-dimensions. This reflects the non-renormalizable
nature of higher-dimensional gauge theories.
Here, we instead see that for $t$ and $u$ channel
interactions between fermions localized at different points, the cross-section
decreases exponentially. The separation acts as a physical 
``point-splitting'' regularization of the non-renormalizable theory, allowing
essentially exact computations for some amplitudes completely independent 
of the physics at the ultimate UV cutoff 
(which is smaller than the separation).

This result, that fermion separation allows us to make unambiguous predictions 
for some quantities in non-renormalizable theories which are 
exponentially insensitive to physics at the cutoff $M_*$, 
is very general. We have already discussed how fermion 
separation provides a physical UV cutoff for the
KK gauge boson exchange. As another example, 
in the context of large extra 
dimensions with low fundamental Planck scale, 
several groups have considered the effects of 
tree-level graviton exchange in the higher dimensions \cite{UVG}. 
For two or more extra dimensions, the sum over the graviton 
KK excitations is UV divergent. Cutting off this divergent 
sum at the scale $M_*$ generates an operator of the form 
\begin{equation}
{\cal O} = \lambda \frac{T_{\mu \nu} T^{\mu \nu}}{M_*^4}
\end{equation}
where $T_{\mu \nu}$ is the 4d energy momentum tensor, and $\lambda$ is an 
unknown constant dependent on the details of how the KK sum is cut off. 
The analysis then proceeds by examining the effect of this particular 
higher-dimension operator on various observables. 
Even if deviations
consistent with this operator are seen experimentally, 
however, this does not provide 
direct evidence for extra dimensions. 
For instance, the operator
may be generated by integrating out a 
single {\it massive} spin 2 particle of mass $\sim M_*$.  
On the other hand, 
if the quarks and leptons are split by some distance $d$, 
the UV divergence 
is automatically cut-off and we can write essentially the exact expression 
for the cross section of e.g., electron proton
scattering including the graviton exchange. 
The expression will only depend on the unknowns $d$, the number of 
extra dimensions $n$ 
and the higher-dimensional Newton constant $G_{N(4+n)}$. 
The only in principle incalculable corrections come from 
the higher-dimensional operators 
suppressed by $M_*$, but these will be suppressed by $\sim e^{-100}$ for the
same reason that proton decay is suppressed to acceptable levels.

It is also important to note that the scattering of split fermions remains 
small even above the scale of quantum gravity $M_*$, say the string scale. 
The reason is still the same; as long as the fermions remain localized 
apart from each other at these
energies, all the new heavy states which come in at $M_*$ 
still need to propagate from one fermion to the other, providing a still 
further suppression of the amplitudes. 
It is interesting that in this scenario, 
we could in principle have the best of all worlds in super-Planckian physics.
The usual expectation is that above $M_*$, all sorts of new physics hit us at
once with a rich and (at least initially) chaotic set of signals. 
We retain this possibility in the 
$s$ channel. But in the $t$ and $u$ channels, 
the interactions between split fermions provide an antiseptic 
environment where the properties of all modes lighter than the inverse fermion
separation (which can include fascinating objects such as bulk gravitons) 
can be unambiguously studied.

Finally, we comment on different possible physics that leads to exponentially
small cross-sections at large $t$, possibly faking our most dramatic signal. 
Consider some 
composite object with some fuzzy size $\Lambda$. For $\sqrt{-t}$ smaller than 
$\Lambda$, we expect that the cross-sections decrease with $\sqrt{-t}$. 
Of course, if these are composite objects like the proton, consisting of
point-like partons, then for $\sqrt{-t} > \Lambda$ we expect the usual 
power-law fall-off with $t$ expected from scattering off the 
point-like constituents, so this can not fake our signal. 
On the other hand, if the fuzziness is like that of a string, 
we may expect that the exponentially decreasing cross-sections persist above
$\Lambda$. However, in this case we do not expect the decreasing amplitude to 
have any simple relationship to the SM amplitudes, whereas for us the 
new cross-section is directly related to the SM as in e.g., 
eqs. (\ref{ep}), (\ref{ee-pol}), (\ref{ee-un}) and 
(\ref{emin-emin}).
This direct correlation between the exponentially falling amplitudes with the 
SM ones is the smoking gun for the observation of fermion separation in extra 
dimensions at future colliders.

%%%%%%%%%%%%%%%%%%%%%%%%
\acknowledgements
%%%%%%%%%%%%%%%%%%%%%%%%
We thank Stan Brodsky, Hooman Davoudiasl, Lance Dixon, Hitoshi Murayama
and Tom Rizzo for useful discussions. N.A.-H. is supported by DOE under
contract DE-AC03-76SF00098 and by NSF under contract PHY-95-14797. Y.G.
and M.S. are supported by the Department of Energy under contract
DE-AC03-76SF00515.

\def\pl#1#2#3{{\it Phys. Lett. }{\bf B#1~}(19#2)~#3}
\def\zp#1#2#3{{\it Z. Phys. }{\bf C#1~}(19#2)~#3}
\def\prl#1#2#3{{\it Phys. Rev. Lett. }{\bf #1~}(19#2)~#3}
\def\rmp#1#2#3{{\it Rev. Mod. Phys. }{\bf #1~}(19#2)~#3}
\def\prep#1#2#3{{\it Phys. Rep. }{\bf #1~}(19#2)~#3}
\def\pr#1#2#3{{\it Phys. Rev. }{\bf D#1~}(19#2)~#3}
\def\np#1#2#3{{\it Nucl. Phys. }{\bf B#1~}(19#2)~#3}

% FIGURE 1
\begin{figure}
\epsfig{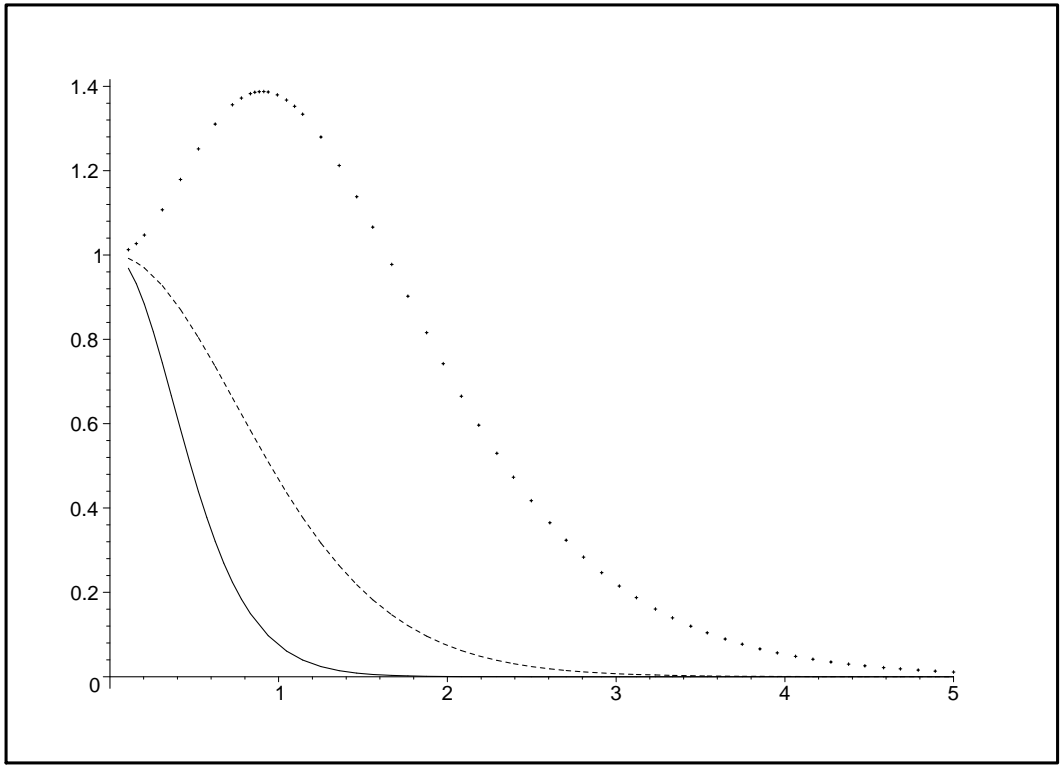}
\caption{$r_\sigma^t$ (the cross section for $t$ channel exchange
in the 5d theory normalized by the corresponding SM cross section)
as a function of $\sqrt{-t}$ in units of TeV. 
We assume $R^{-1}=1\,$TeV. The dotted, dashed and solid curves are for
separations of $d/R=1$, $\pi/2$ and $\pi$ respectively.}
\end{figure}

% FIGURE 2
\begin{figure}
\epsfig{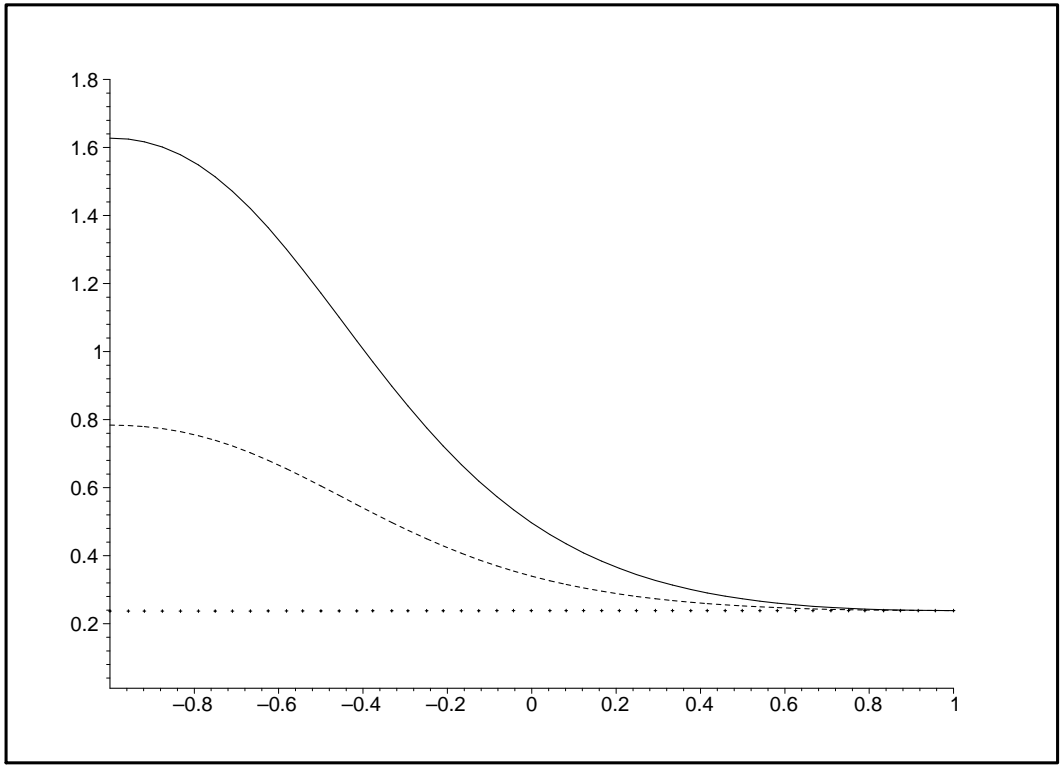}
\caption{$r_\sigma^s$ (the cross section for $s$ channel exchange, e.g.\ 
$e^+ e^- \to \mu^+ \mu^-$,
in the 5d theory normalized by the corresponding SM cross section)
as a function of the scattering angle, $\cos\theta$. 
We assume $R^{-1}=4\,$TeV and $\sqrt{s}=1.5\,$TeV. 
The dotted, dashed and solid curves are for
separation of $d/R=0$, $1$ and $\pi$ respectively.}
\end{figure}

% FIGURE 3
\begin{figure}
\epsfig{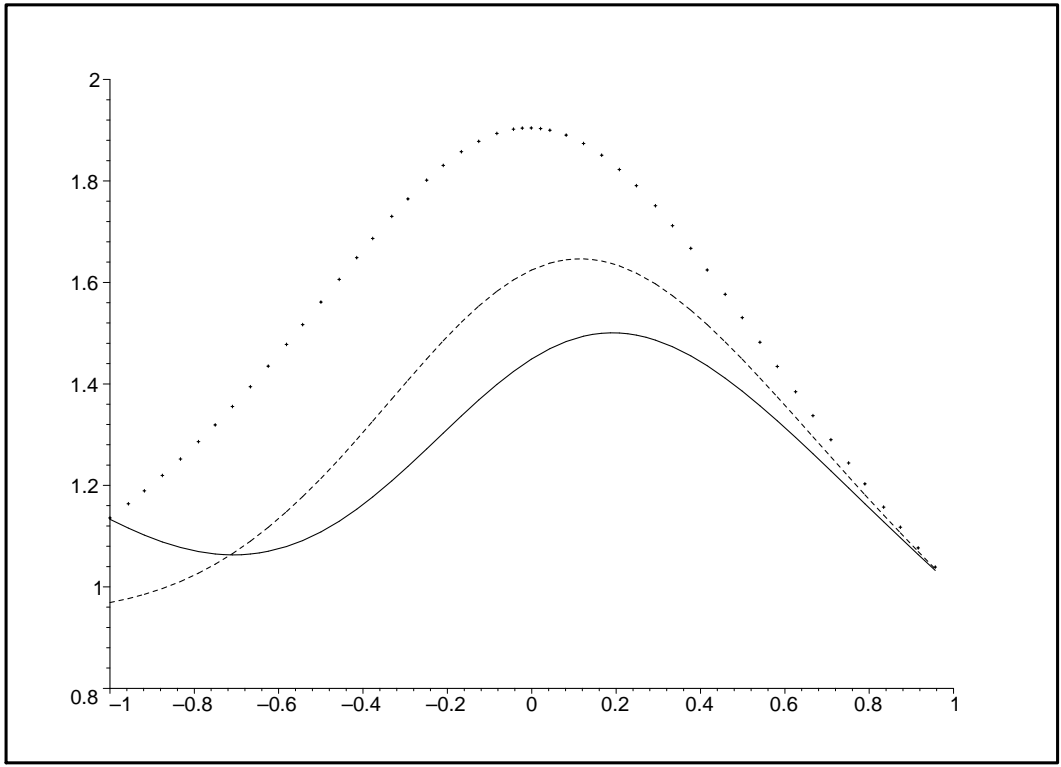}
\caption{Same as fig.~2 for $r_\sigma^{st}$
(the cross section for $s$ and $t$ channel exchange, e.g.\ 
$e^+ e^- \to e^+ e^-$,
in the 5d theory normalized by the corresponding SM cross section).}
\end{figure}

\end{document}